\documentclass{aa}  
\usepackage{graphicx}
\usepackage{natbib}
\usepackage{ulem}
\usepackage{txfonts}
\begin{document}
   \title{Near-IR internetwork spectro-polarimetry at different heliocentric angles}

   \subtitle{}

   \author{M. J. Mart\' inez Gonz\'alez\inst{1}
          \and
          A. Asensio Ramos\inst{2}
	  \and 
	  A. L\'opez Ariste\inst{3}
	  \and 
	  R. Manso Sainz\inst{2}}

   \offprints{M. J. Mart\' inez Gonz\'alez}

   \institute{LERMA, Observatoire de Paris-Meudon, 5 place de Jules Janssen, 92195, Meudon, France\\
              \email{Marian.Martinez@obspm.fr}
         \and
             Instituto de Astrof\' isica de Canarias, V\' ia L\'actea S/N, 31200, La Laguna, Spain\\
             \email{aasensio@iac.es}
         \and
             THEMIS-CNRS UPS 853, V\' ia L\'actea S/N, 31200, La Laguna, Spain\\
             \email{arturo@themis.iac.es}}

   \date{Received ; Accepted }

 
  \abstract
   {}
   {The analysis of near infrared spectropolarimetric data at the internetwork at different regions on the solar surface could offer constraints to reject current modeling of these quiet areas.}
   {We present spectro-polarimetric observations of very quiet regions for different values of the heliocentric angle for the Fe\,{\sc i} lines at $1.56$ $\mu$m, from disc centre to positions close to the limb. The spatial resolution of the data is $0.7-1''$. We analyze direct observable properties of the Stokes profiles as the amplitude of circular and linear polarization as well as the total degree of polarization. Also the area and amplitude asymmetries are studied.}
   {We do not find any significant variation of the properties of the polarimetric signals with the heliocentric angle. This means that the magnetism of the solar internetwork remains the same regardless of  the position on the solar disc. This observational fact discards the possibility of modeling the internetwork as a Network-like scenario. The magnetic elements of internetwork areas seem to be isotropically distributed when observed at our spatial resolution.}
   {}


   \maketitle
%

\section{Introduction}

The presence of magnetic fields in the solar internetwork was discovered more than 30 years ago \citep{livingston_75,smithson_75}. Since that work, improvements on the instrumentation have allowed the observation of the full Stokes vector in those regions with a signal-to-noise ratio good enough to retrieve the magnetic field vector from the observational data. Nevertheless, the internetwork magnetic topology remains yet very  indistinct and shows a growing  complexity as we improve the quality of the data. \cite{martin_87} observed that in her videomagnetographs at a spatial resolution of $3''$ the longitudinal component of internetwork magnetic fields was present everywhere in the solar disc. She immediately concluded that these magnetic structures should have very small scales and a very tangled geometry, giving as an example a scenario in which the magnetic fields in the internetwork would consist of a maze of small loops.

As spectro-polarimeters have made possible the precise detection of these signals, the efforts have concentrated on the study of the distribution of magnetic fields at disc centre. Different methodologies (mainly based upon the use of the Zeeman and Hanle effects) shed light on  different aspects of the internetwork magnetism. Concerning the Zeeman effect, the Stokes $Q$, $U$ and $V$ signals detected on the most widely used spectral lines (Fe\,{\sc i} at $1.5$ $\mu$m and 630 nm) are very weak at the best spatial resolutions of 0.5-1$''$. Moreover, the Stokes $I$ profile seems to come from a field free atmosphere, as expected from weakly polarized media \citep{jorge_99}. The filling factor of the magnetic elements retrieved from the analysis of these signals is always around 2 \% \citep{khomenko_03, jorge_ita_03, marian_spw4}. The rest of the resolution element would be filled with very weak magnetic fields. Another possibility is that the real element is filled with mixed polarity magnetic fields which would partially cancel out due to the lack of spatial resolution. The magnetic field strength distributions recovered for this 2 \% of the resolution element appear to show a preference for magnetic fields around the equipartition field at photospheric heights and even weaker \citep{khomenko_03, marian_spw4, julio07}. Additionally, \cite{andres_07} have shown the first direct observational evidence of flux cancellation in the internetwork (note that their internetwork region is surrounded by a very enhanced network structure), showing that more than 95 \% of the magnetic flux is cancelled in the resolution element ($\sim 1''$). This means that the above-mentioned distributions could not give a complete vision of the internetwork magnetism. \cite{andres_07} give an amount of $250$ G for the mean magnetic field in the resolution element (note that the values of the magnetic flux density obtained by means of the Zeeman effect are below 10 Mx/cm$^2$). This would mean that the magnetism of the internetwork could play an important role on the solar global magnetism. This has also been pointed out by works using the Hanle effect \citep{javier_04}.

The study of regions in different positions of the solar disc could represent a strong constraint to reject models for the solar internetwork. Using the 630 nm lines with a spatial resolution of $\sim 1''$, \cite{lites_02} built a histogram of both circular and linear polarization signals in two quiet regions, one at disc centre and another at $\mu=0.82$ (being $\mu$ the cosine of the heliocentric angle). He did not observe significant linear polarization signals in neither of the two regions.  However, Figure 9 in \cite{lites_02} shows that the histograms of circular polarization do not present any variation for those signals whose integrated signal is below 0.005. \cite{meunier_98}, studying integrated polarization, and  \cite{harvey_07} using magnetograms, show the presence of a horizontal component of the magnetic field everywhere in the solar disc.
In this paper we present the first study of the internetwork at several positions on the solar disc using high quality $0.8''$ spectro-polarimetric data.

\section{Observations and data reduction}

The observations consist of 2-dimensional maps of the very quiet Sun taken at different positions covering the two solar
hemispheres. In each pixel of these maps we recorded the four Stokes parameters of the Fe\,{\sc i} lines at 1.56 $\mu$m. The observations were performed at the Vacuum Tower Telescope (VTT, Observatorio del Teide) during August 2000 and July 2006 with the Tenerife Infrared Polarimeter (TIP) instrument \citep{manolo99}. The observations during 2006 made use of the adaptive optics system 
attached to the telescope. This led to a much improved spatial resolution and to an important decrease in the exposure time required to
reach a noise level comparable to those obtained for the data of August 2000. The maps of August 2000 cover the following positions (we also indicate the quadrant of each
scan): $\mu$=0.88 (N), $\mu$=0.4 (N) and $\mu$=0.28 (E). The maps of July 2006 were taken at $\mu$=1 (disc centre), $\mu$=0.81 (SW), $\mu$=0.73 (S), $\mu$=0.62 (W) and $\mu$=0.46 (W). Active regions or very enhanced Network were avoided using the Ca\,{\sc II} K filter available simultaneously to the observations, since bright regions in this filter are good indicators of magnetic flux concentrations \citep{lites_99}.

The data reduction consisted in the subtraction of dark current, flatfield correction and demodulation of the images. After that, we also corrected for other residual patterns, mostly polarization-dependent fringes \citep{semel_03}. Most of the instrumental crosstalk can be removed from the data using the calibration optics located before the beam splitter \citep{schlichenmaier_02}. However, the coelostat configuration of the telescope has to be modeled \citep{manolo99}. The residual crosstalk from Stokes $I$ to Stokes $Q$, $U$ and $V$ was removed by forcing the continuum of the polarization profiles to zero. The residual crosstalk between Stokes $Q$, $U$ and $V$ is very difficult to remove and a few percent may still remain. The last step of the data reduction was a de-noising procedure based on Principal Component Analysis. The noise level in the polarization profiles of the 2006 data sets is in the range $3\times 10^{-5}$ - $8\times 10^{-5}$ in units of the mean continuum intensity, I$_\mathrm{c}$. The value in the 2000 data is $\sim 10^{-4}$ I$_\mathrm{c}$. This means that the signal to noise ratio is $\sim 15$ (taking the most probable polarization amplitude as $8\times 10^{-4}$ I$_\mathrm{c}$) and $\sim 8$ in the 2006 and 2000 data sets, respectively. The spatial resolution is $\sim 1''$ for the 2000 data and $0.7-0.8''$ for the 2006 observations.

In order to analyze the data in a reliable way, we select those profiles which have a degree of polarization $A_P$ (maximum of $\sqrt{Q^2+U^2+V^2}$) higher than $4\times 10^{-4}$ I$_\mathrm{c}$. Following this criterion we do not introduce much bias in the results since we select all kind of magnetic field vector configurations. In the disc centre observations we have detected some small accumulations of high Stokes $V$ amplitude ($> 0.01$ I$_\mathrm{c}$) that may correspond to Network points. Far from the disc centre one expects  Network fields to be more inclined. Then, we reject those points with amplitudes of total polarization higher than $> 0.01$ I$_\mathrm{c}$. This criteria is not perfect since linear polarization does not grow as fast as circular polarization decreases when the magnetic field inclination with respect to the line-of-sight increases. Anyway, these points represent at most 1 \% of the field of view. The selected internetwork profiles account for at least  $95$ \% of the observed field of view in all maps.

\section{Analysis of circular and linear polarization}

From the selected profiles we compute the amplitude of the circular polarization ($A_V$) in the following way. We find the position of the maximum and the minimum of the Stokes V profile. We compute the mean value of the profile around these positions (we take 3 points before and after the maximals). The amplitude $A_V$ is the mean value of the absolute value of the amplitude of the two lobes. The linear polarization ($A_L$) is defined as the mean value around the maximum of $\sqrt{Q^2+U^2}$. The same calculation is performed to obtain the total polarization $A_P$. Both the $1.5648$ and $1.5652$ $\mu$m spectral lines show the same behaviour and we present here results for just the \hbox{$1.5648$ $\mu$m} line whose larger amplitudes in the Stokes parameters makes the results less subject to noise effects.

\begin{figure}
\includegraphics[width=9cm]{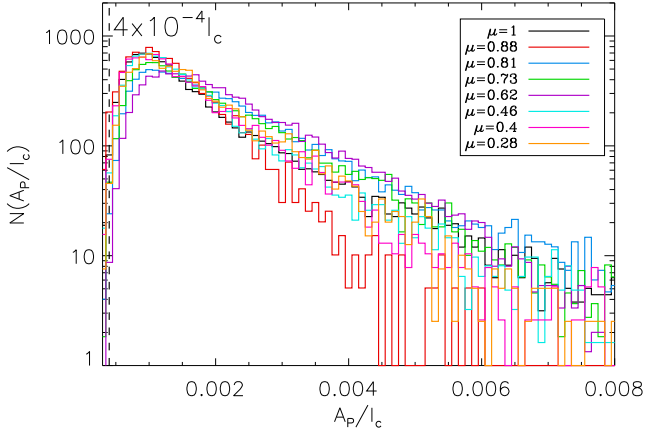}
\caption{Histograms of the total polarization $A_P$ for all the observed areas on August 2000 (those corresponding to $\mu=0.88, 0.4, 0.28$) and July 2006 (those corresponding to $\mu=0.81, 0.73, 0.62, 0.46$). The improvement of the spatial resolution from 1 to 0.7-0.8$''$ has made it possible to detect a large amount of very weak signals. The results show that there is no variation of the signal for different heliocentric angles. The vertical dashed line
indicates the chosen detection limit.}
\label{ampl_pol}
\end{figure}

The comparison of the histograms of the polarization amplitudes between the data of 2000 and that of 2006 proved to be very difficult. First, the different spatial and spectral resolutions (the sampling in wavelength in 2000 data was twice the one of the 2006 data) and different noise levels result in a different polarimetric sensitivity. Second, both observations have different number of pixels. This can be partially solved by normalizing all the histograms to its area. However, the fact that both data sets have different polarimetric sensitivities makes its comparison impossible after the normalization. Figure \ref{ampl_pol}  shows the degree of polarization ($A_P$) of the two different data sets. There is no evident difference between the signals recorded either on 2000 or on 2006. However, for the sake of having reliable conclusions we only explicitly compare the histograms for the 2006 data. In order to compare all the observations we analyze other properties of the histograms that do not depend on the normalization.

\begin{figure}
\includegraphics[width=9cm]{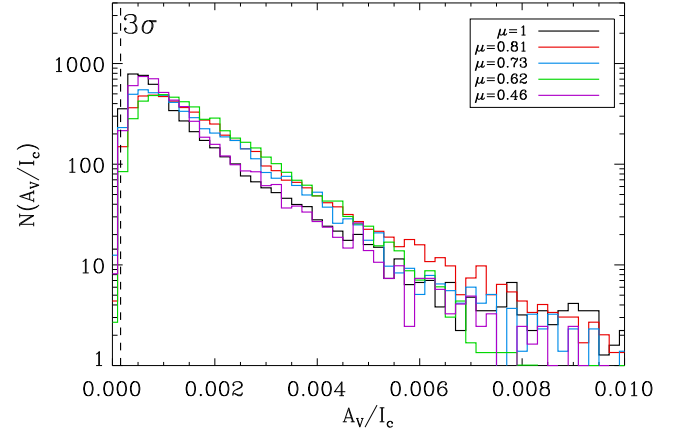}\\
\includegraphics[width=9cm]{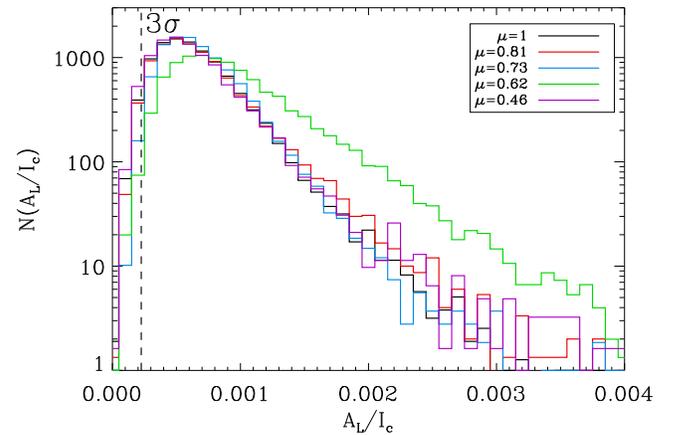}\\
\includegraphics[width=9cm]{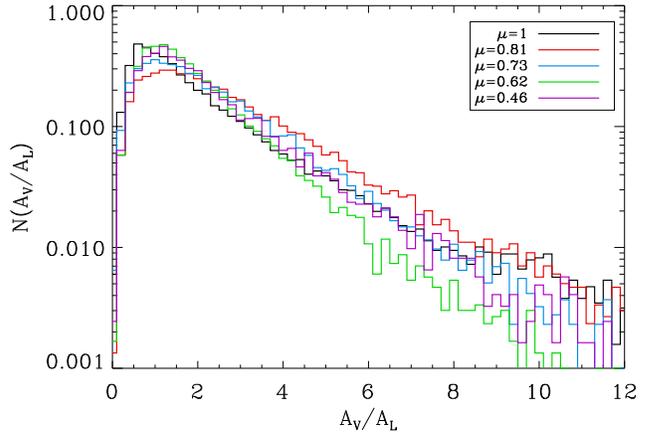}\\
\caption{Histograms of circular polarization (upper panel), linear polarization (central panel) and its ratio (bottom panel) for all the observed areas on July 2006. There is no
variation of the distributions with heliocentric angle.}
\label{ampl_circ_lineal}
\end{figure}

The top and center panels of Fig. \ref{ampl_circ_lineal} show the histograms of the circular and linear polarization amplitudes of the $1.5648$ $\mu$m line for all the observed positions on 2006. We have normalized the $A_V$ and $A_L$ values to the mean continuum intensity of each map to avoid the effect of limb darkening. There is no evident variation of both circular or linear polarization signals in all the studied positions. However, the histogram of linear polarization at $\mu$=0.62 is different from the others. A visual inspection of the data reveals that this data set is highly contaminated by cross-talk from Stokes $V$ to $Q$ and $U$. We conclude that, irrespective of the position on the solar disc, the polarimetric signals on the internetwork seem to be the same at our $0.7-0.8''$ spatial resolution. The bottom panel of Fig. \ref{ampl_circ_lineal} shows that the ratio between the circular and linear polarization does not change significantly over the solar disc. Since the signal to noise ratio is at least 15, we do not expect the results to be dominated by noise effects so that we can affirm that the observed pattern is of solar origin.

According to the bottom panel of Fig. \ref{ampl_circ_lineal}, the most probable value for the ratio between circular and linear polarization is close to 1. This can be a surprising result if we have in mind all the results found in the literature concerning the internetwork magnetism using the $630$ nm lines \citep[e.g.,][]{lites_02}. At this spectral range no linear polarization can be found at 0.5-1$''$ spatial resolution \citep{lites_02, marian_tesis_07}. However, the linear polarization sensitivity of the $1.5$ $\mu$m pair of lines is very different from that of the 630 nm pair. \cite{egidio} define some useful magnitudes that can be used to get an idea of the sensitivity of a spectral line to the circular and linear polarization. They are defined as:
\begin{eqnarray}
s_V&=&\frac{\lambda}{\lambda_{ref}}\overline{g}d_c  \nonumber \\
s_L&=&(\frac{\lambda}{\lambda_{ref}})^2 \overline{G} d_c,
\end{eqnarray}
where $s_V$ is the circular polarization sensitivity index and $s_L$ is the corresponding linear one, while $\lambda_{ref}$ is a fixed reference wavelength. Consequently, these numbers have only sense when comparing two spectral lines. The symbol $\overline{g}$ stands for the effective Land\'e factor of the transition and $d_c$ is the central depression in terms of the continuum intensity, defined as:
\begin{equation}
d_c=\frac{I_c-I(\lambda_0)}{I_c},
\end{equation}
being $I(\lambda_0)$ the intensity at the line core. The symbol $\overline{G}$ plays the same role as the effective Land\'e factor but for the linear polarization and it is defined as:
\begin{equation}
\overline{G}=\overline{g}^2-\delta,
\end{equation}
where $\delta$ is a quantity that depends on the quantum numbers of the transition \cite[see][]{egidio}. Both $1.5648$ $\mu$ and $630.2$ nm spectral lines  have $\delta=0$. The ratios between the circular and linear polarization indices of the infrared and visible spectral lines are:
\begin{eqnarray}
\frac{(s_V)_{1.5 \mu m}}{(s_V)_{630 nm}}&=&1.37 \nonumber \\
\frac{(s_L)_{1.5 \mu m}}{(s_L)_{630 nm}}&=&4.17.
\end{eqnarray}
This means that the circular polarization sensitivity is similar in both spectral ranges but the infrared lines present a linear polarization sensitivity that is 4 times larger
than the visible lines. This fact explains the observed behaviour of the presence of ubiquitous linear polarization signals in the infrared and the lack of them in the visible for similar spatial resolutions.

\begin{figure*}
\includegraphics[width=9cm]{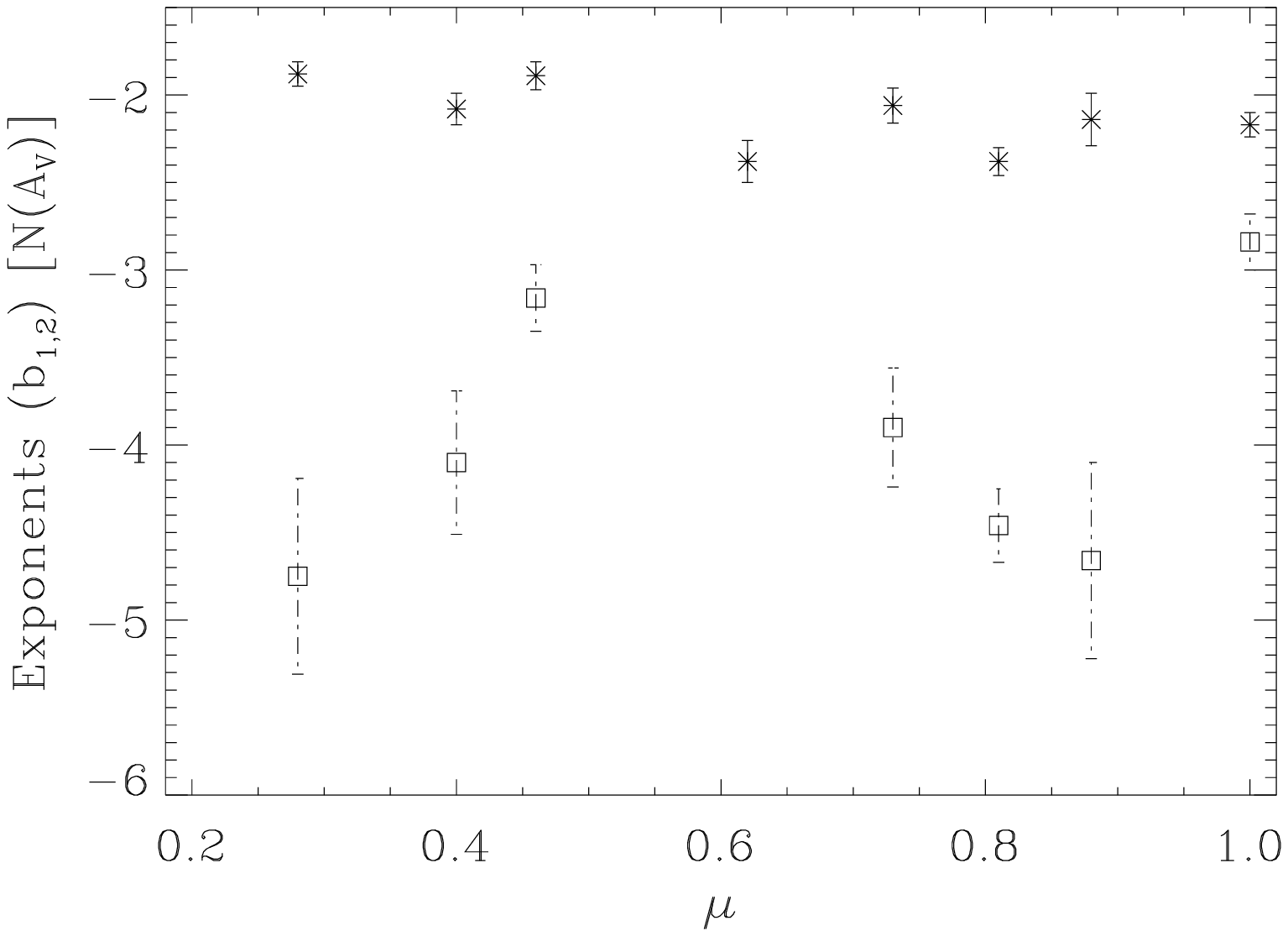}
\includegraphics[width=9cm]{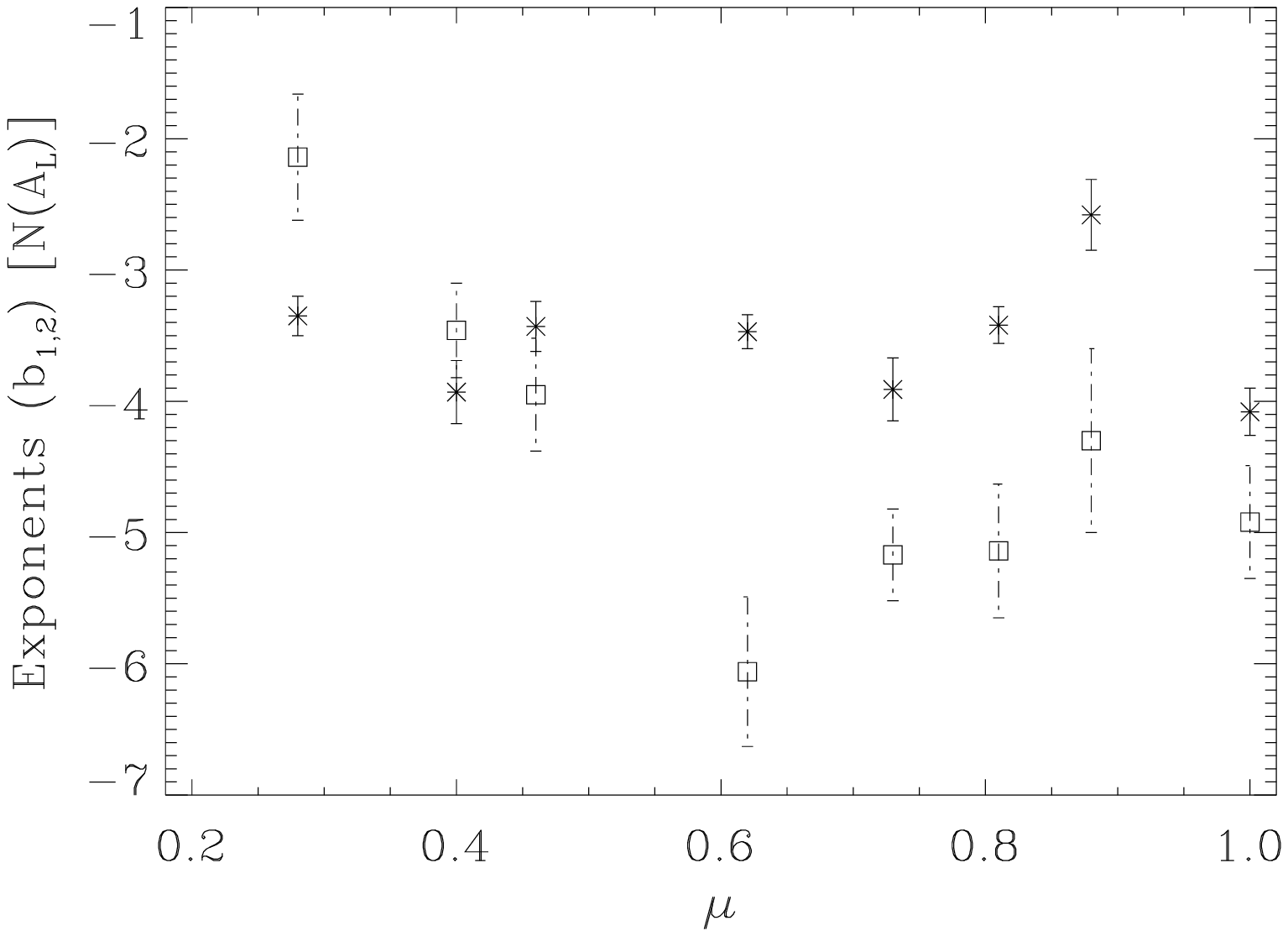}\\
\includegraphics[width=9cm]{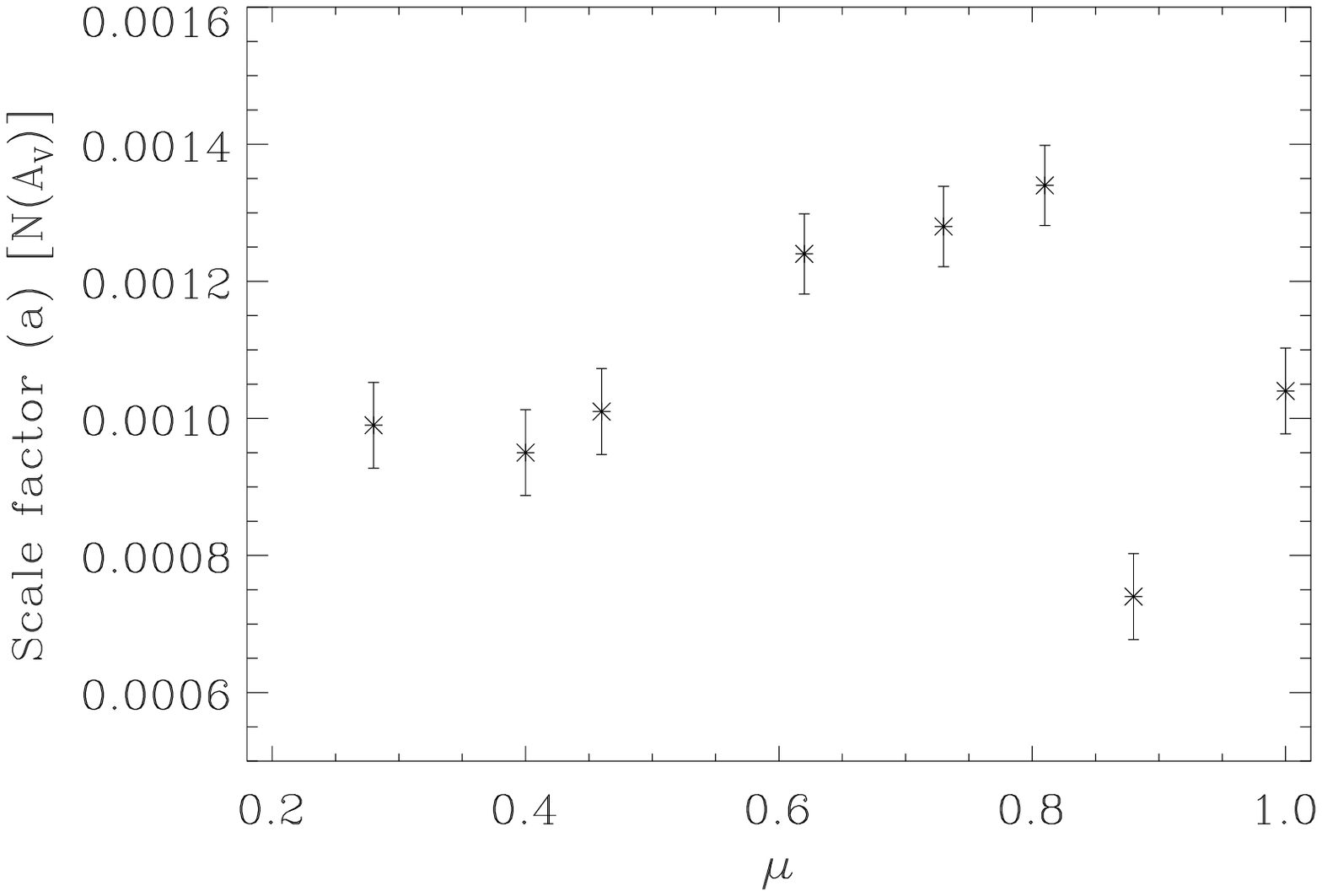}
\includegraphics[width=9cm]{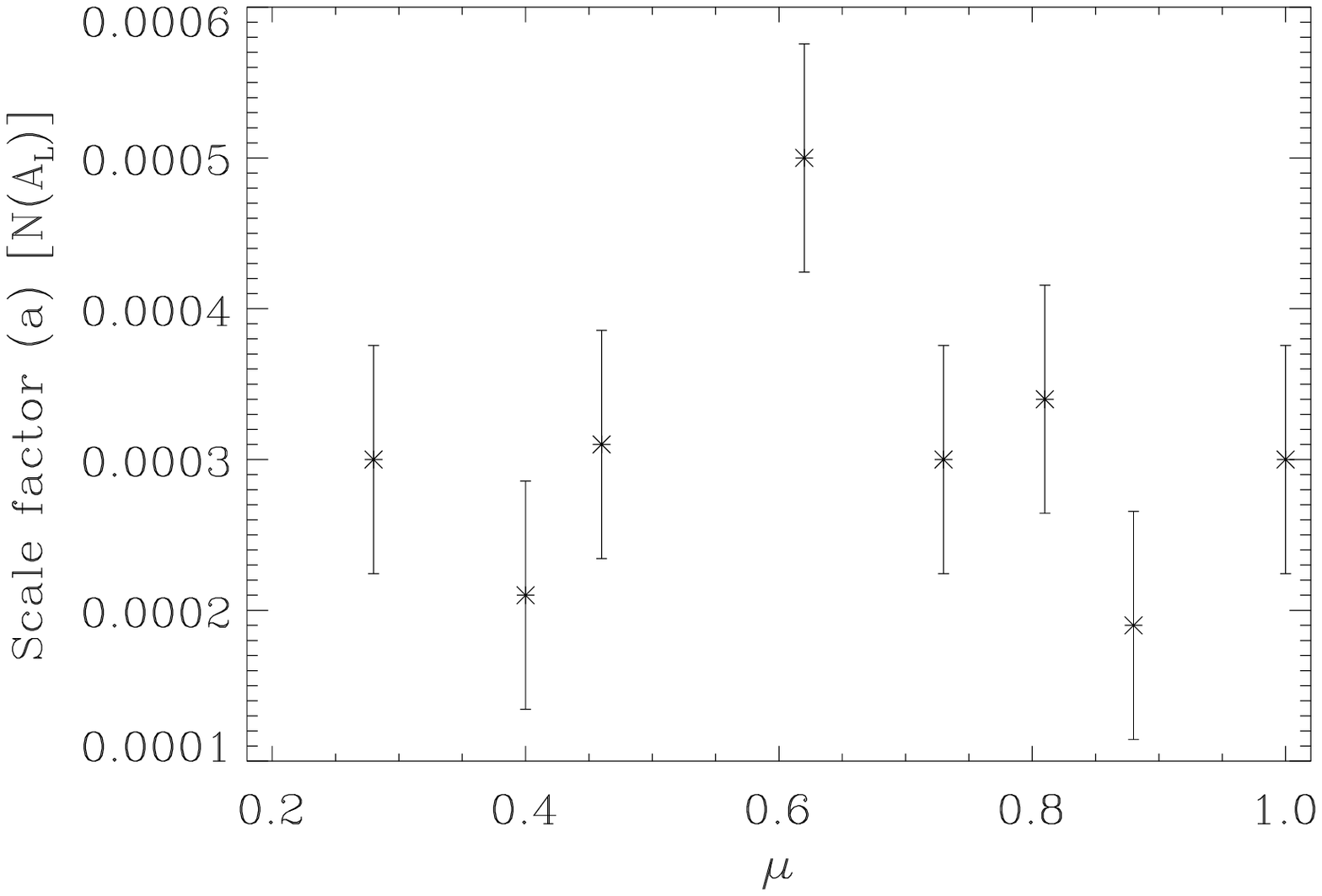}
\caption{Upper panels: exponents of the two power laws ($b_1$ shown in stars and $b_2$ shown in squares) that fit the tails of the distributions of circular and linear amplitudes shown in Fig. \ref{ampl_circ_lineal}. Bottom panels: values of scale factors $a$ for the tails of the distributions.}
\label{param}
\end{figure*}

All the histograms of Figs. \ref{ampl_pol} and \ref{ampl_circ_lineal} (and also the circular and linear polarization histograms of year 2000) present a common behaviour: a peak is found for values markedly above the detection limit. In the case of the total polarization, we can see that the maximum of all histograms is far from the chosen noise threshold of $4\times 10^{-4}$ I$_c$. Also the drop towards small amplitudes in the case of the circular and linear polarization occurs for values larger than three times their noise level. Such behaviour can be expected from cancellations of magnetic fields due to a decrease of the spatial resolution. In any case we do not discard a bias introduced by noise or other systemathic deffects (e. g. interference fringes) in the determination of the amplitudes. This issue is now under investigation.

In order to further compare the whole data sample we choose two different functional forms to fit the tails of the histograms. First, a combination of two power laws ($y \propto x^{b_i}$). The first one (parameterized by $b_1$) describes the drop from the peak to $\sim 7\times 10^{-3}$ I$_\mathrm{c}$ for circular polarization or to $\sim 3\times 10^{-3}$ I$_\mathrm{c}$ for linear polarization, while the second one (parameterized by $b_2$) describes the drop for the rest of the tail above these points. Second, a decreasing exponential law ($y \propto \exp[-x/a]$) starting from the peak of the histograms. This second option is chosen following previous works on quiet Sun magnetism in the infrared. These parameters ($a$ and $b_i$) are independent of the normalization of the histograms and they are useful to compare the behaviour of these histograms with the heliocentric angle.


Figure \ref{param} shows the dependence of these three parameters with the heliocentric angle. Error bars represent statistical uncertainty. The left panels present the results for the circular polarization amplitudes while the right panels refer to the linear polarization amplitudes. No clear trend is found. Perhaps, it could be possible to identify a trend in the value of $b_2$ with $\mu$, indicating the presence of magnetic fields with a preferential orientation. In any case, the poor statistics on the tails make this conclusion delicate. We cannot be sure if this behaviour is due to some isolated Network patches, high magnetic flux pixels that may contaminate the histograms or just statistical noise. In any case they represent a small percentage of the points under study. The key fact is that the majority of the signals (coming from internetwork areas) have no trend with the heliocentric angle. This analysis states that all the positions on the internetwork quiet Sun are indistinguishable from the point of view of the distribution of polarization signals. Our data at $0.8-1"$ reveals no clear variation of the internetwork magnetism with the heliocentric angle, supporting the idea of an isotropically distributed field.


\begin{figure*}
\includegraphics[width=9cm]{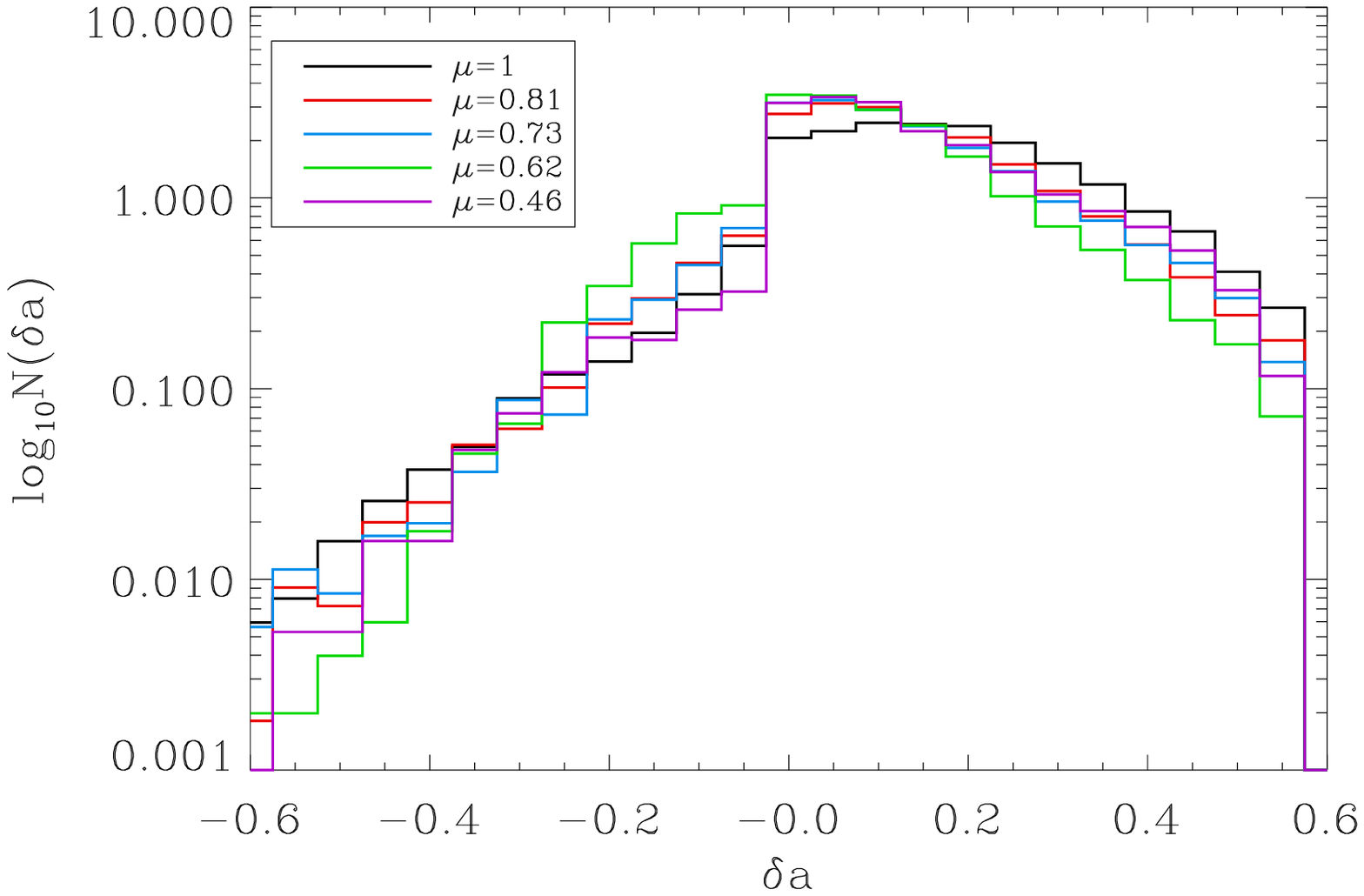}
\includegraphics[width=9cm]{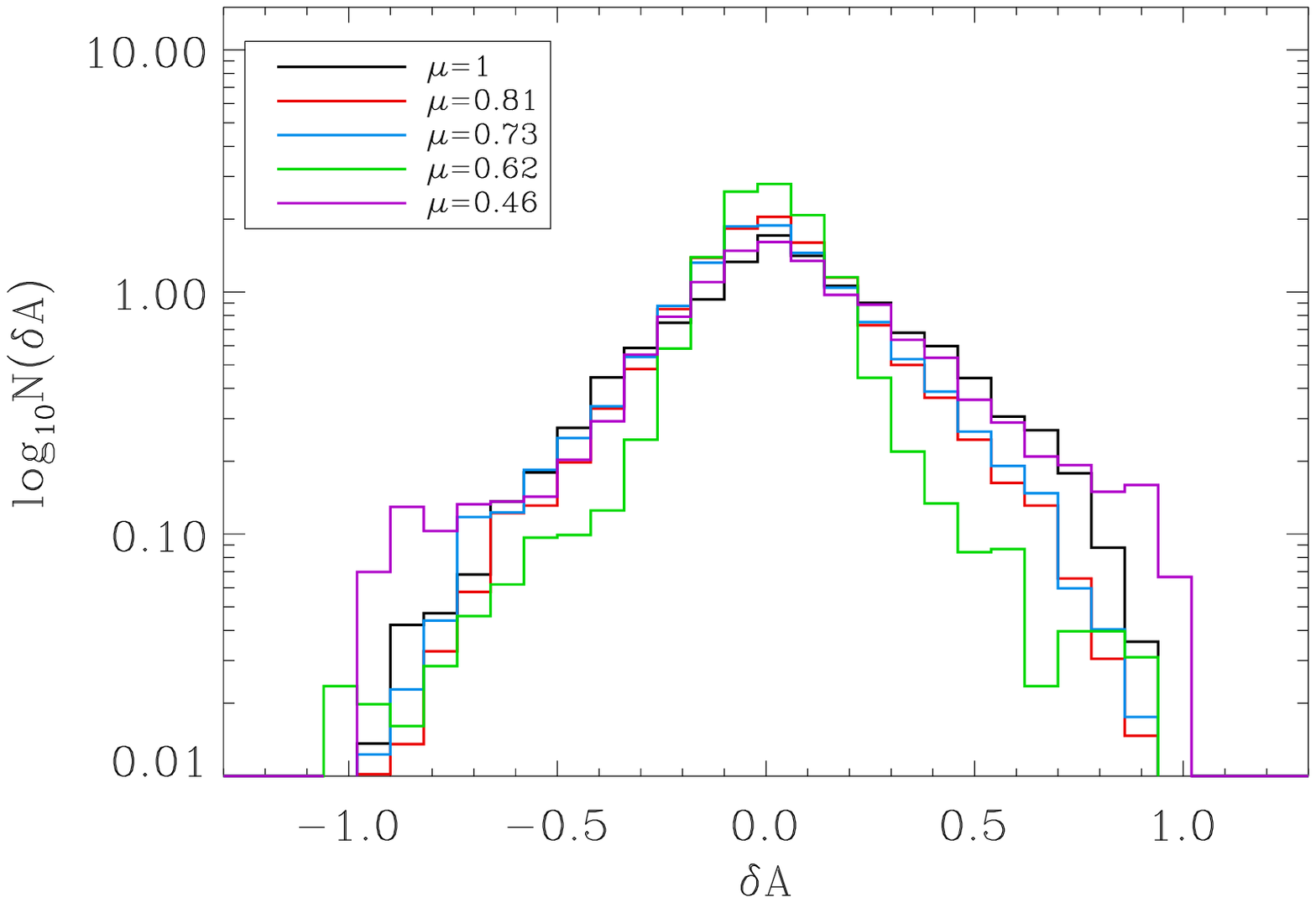}
\caption{Histograms of the amplitude (left panel) and area (right panel) asymmetries for the regular Stokes V profiles.}
\label{asim_circ}
\end{figure*}

\begin{figure*}
\includegraphics[width=9cm]{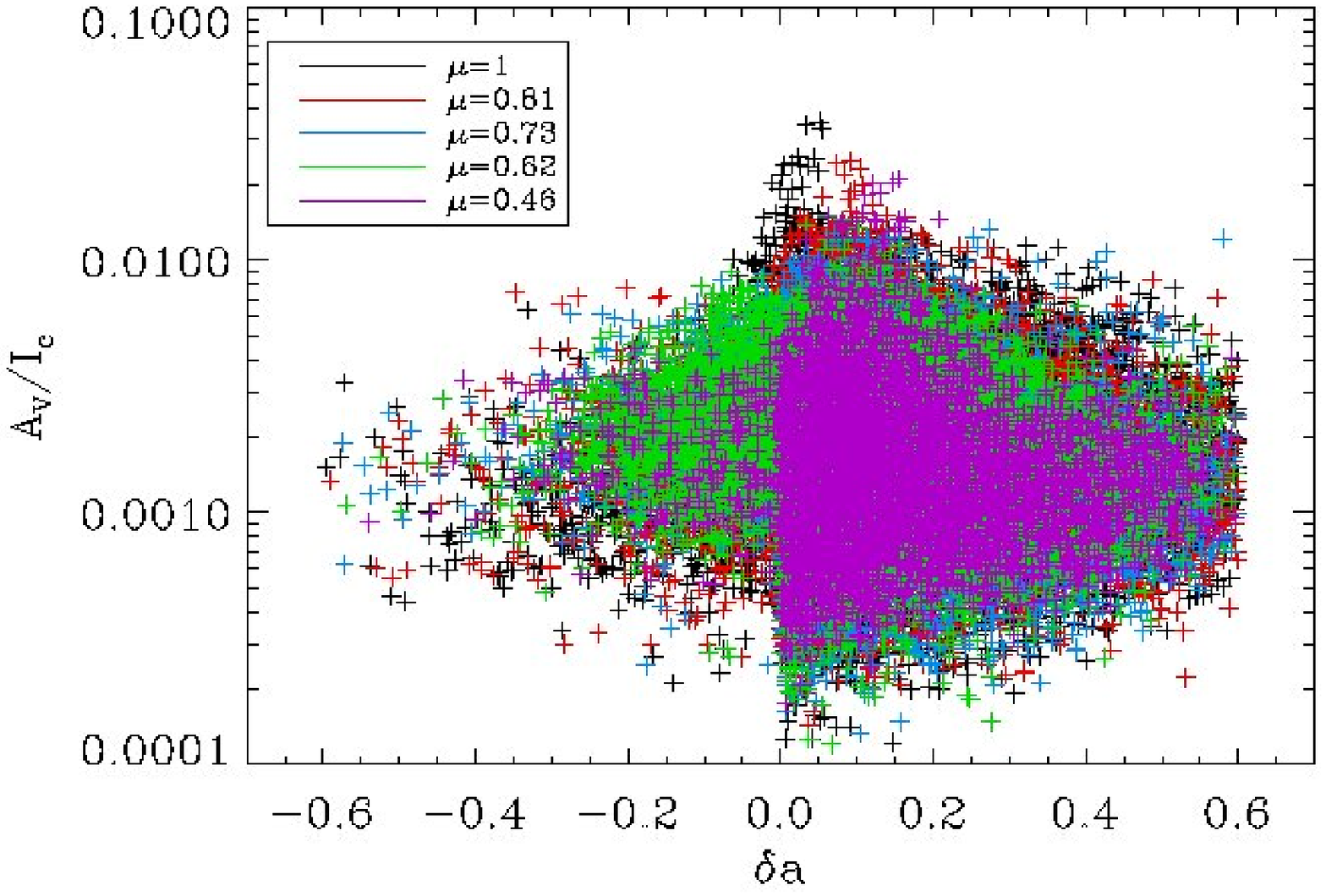}
\includegraphics[width=9cm]{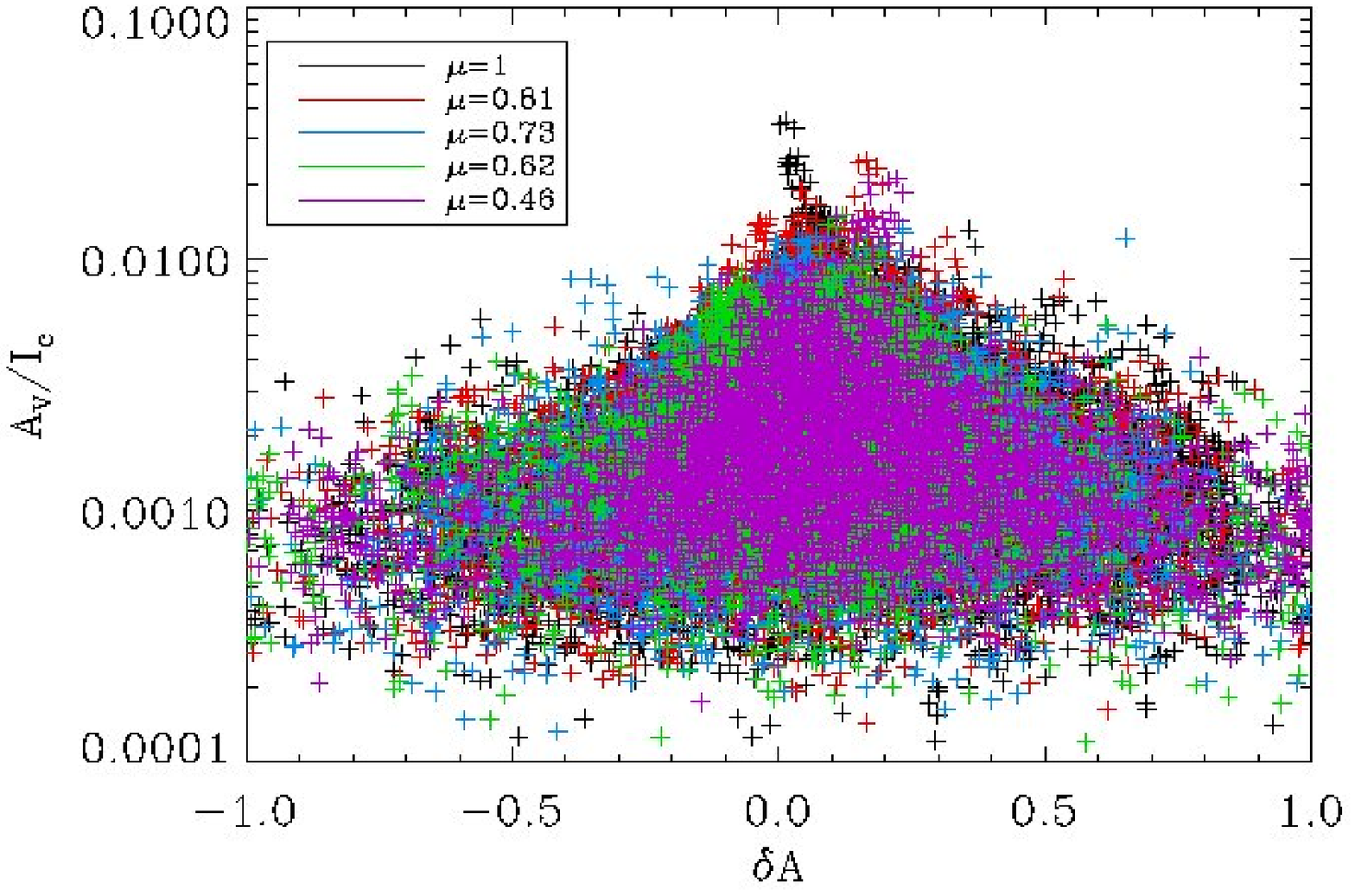}
\caption{Amplitude (left panel) and area (right panel) asymmetries plotted versus the amplitude of the regular Stokes V profiles.}
\label{asim_vs_circ}
\end{figure*}

\section{Analysis of amplitude asymmetries of the Stokes V profiles}

For the computation of the Stokes V asymmetries we consider only the regular (two-lobbed) profiles. In order to select them, we first determine the number of lobes (and their position in wavelength and amplitudes) of both the $1.5648$ and $1.5652$ $\mu$m lines. Due to small residual patterns of the reduction, we can find some profiles where one of the spectral lines presents two lobes and the other one presents three. In order to extract the real lobes (whose number should be the same in both lines) we impose both spectral lines to have zero--crossings compatible with an error of $0.6$ km/s. If the line has two lobes, the zero-crossing is defined as the point where the Stokes V line profile reaches zero in between the lobes. If it presents three lobes there will be two zero crossings: one between the red and the centre lobes and a second one between the blue and centre ones. Finally, we only select the two lobes that are compatible. This data selection reduces the sample of ``regular'' profiles to 63$\pm 3$ \% of the field of view, which still represents a large fraction and the results will have a strong statistical significance.

Figure \ref{asim_circ} shows the amplitude asymmetry of the regular Stokes V profiles, defined as \cite{sami_84}:
\begin{equation}
\delta a=\frac{a_b-a_r}{a_b+a_r},
\end{equation}
where $\delta a$ accounts for the amplitude asymmetry while $a_b$ and $a_r$ refer to the absolute values of the blue and red lobes of the Stokes V profile, respectively. The area asymmetry is computed as:
\begin{equation}
\delta A=\frac{A_b-A_r}{A_b+A_r},
\end{equation}
being $\delta A$ the area asymmetry while $A_b$ and $A_r$ are the absolute values of the areas of the blue and red lobes, respectively. Computing the area asymmetry implies choosing the boundaries for the integration of each lobe. In our case the integration interval is different for each profile. We select the initial point for the integration of the blue lobe as the position on the blue wing closest to the zero crossing which has a value lower than 3$\times 10^{-5}$ I$_c$ while the ending point is chosen to be the zero crossing of the profile. This last one is also the lower limit for the integration of the red lobe and the final one is also the closest point to the profile with a value lower than 3$\times 10^{-5}$ I$_c$ but in the red wing.

As shown in Fig. \ref{asim_circ}, the amplitude and area asymmetries do not change significantly in all the studied positions. Small variations can be detected but no clear trend with $\mu$ is found. The results for the amplitude asymmetry are compatible with those obtained by \cite{khomenko_03} at disc centre, including the upper limit of 0.6 for the positive values. For the area asymmetry we found a wider range of values even though the distributions are symmetric and centered at $\delta A=0$, also consistent with
what is found by \cite{khomenko_03}.

Figure \ref{asim_vs_circ} shows the Stokes V amplitude versus the amplitude (left panel) and area asymmetries (right panel). Left panel shows that negative amplitude asymmetries appear to present amplitudes preferentially in the range between $3\times 10^{-4}$ and $4\times 10^{-3}$ I$_\mathrm{c}$. For the positive asymmetries the case is more complicated, where one is able to distinguish at least three different regimes. Like for the case of negative asymmetries, Stokes $V$ amplitudes in the range $0.3-4\times 10^{-3}$ I$_\mathrm{c}$ seem to cover the whole range of values (up to an upper limit of 0.6). For higher amplitudes, the higher the value of $A_V$ the smaller the value of $\delta a$. Interestingly, amplitudes lower than $\sim 3\times 10^{-4}$ I$_\mathrm{c}$ show the inverse case: the higher the amplitude the higher the asymmetry.

\section{Discussion and conclusions}

We have analyzed high quality spectro-polarimetric  data of the Fe\,{\sc i} lines at $1.56$ $\mu$m in order to infer information about  the internetwork magnetism at different positions on the Sun's surface. The whole field of view presents significant signal, meaning that the magnetic fields pervade the observed areas.

We have found that the circular and linear polarization amplitudes do not have any clear dependence on the heliocentric angle. This fact goes against a Network-like scenario for the internetwork: quasi--vertical flux tubes cannot explain this observational result, nor in fact any field topology with a preferred orientation within the field-of-view. An isotropical distribution of magnetic fields, oriented in all directions in the whole field of view, is on the other hand expected to show this behaviour. \cite{marian_07} found that at least 10-20 \% of the magnetic flux in the internetwork is connected by low-lying loops. Consequently, the scenario proposed by \cite{martin_87} of an internetwork characterized by a myriad of small loops is a very reasonably idea that is compatible with all the observational constraints presented in this work. Of course one can think of other scenarios that are compatible with the observations. \citep{stenflo_87, rafa_04, javier_04} adopt turbulent internetwork magnetism, \cite{jorge_00} proposes a micro-structuration of the atmosphere (MISMA) to explain all the magnetic phenomena on the solar surface. All of them are compatible with the presented results and our efforts should be headed towards finding more constraints to reject some of them and strengthen others.

%

The size of the magnetic structures can also be constrained by the information presented in this study. First, improving the spatial resolution, we do not see a global increase in the signals. \cite{lites_04} found no increment of the magnetic flux density from $1"$ to $0.6"$. Recently \cite{lites_07} computed a magnetic flux density of about $11$ Mx/cm$^2$ at 0.3$''$ using HINODE's data, which is compatible with the value of $10$ Mx/cm$^2$ found by \cite{martin_87} at $3''$. This means that, either the magnetic field structures are already resolved at $\sim 0.5''$ or we are very far from resolving them. The fact that the polarimetric signals do not vary along the solar surface would point towards very small structures as the responsibles for the internetwork magnetism. The size of these magnetic structures is something not yet constrained by the observations.

We have presented a study of high quality spectro-polarimetric data in different positions of the solar surface, from the disc centre towards $\mu=0.28$. This is the first step of the study of the variation on the magnetism of the internetwork with the heliocentric angle. Much more work has to be done by retrieving physical parameters as the magnetic field strength vector, magnetic flux, etc. to really constrain the modeling of the internetwork and reject models that are not compatible with the results.

\begin{acknowledgements}
This article is based on observations taken withthe VTT telescope operated on the island of Tenerife by the Kiepenheuer-Institut f\"ur Sonnenphysik in the Spanish Observatorio del Teide of the Instituto de Astrof\'{\i}sica de Canarias (IAC). 
\end{acknowledgements}


\begin{thebibliography}{26}
\expandafter\ifx\csname natexlab\endcsname\relax\def\natexlab#1{#1}\fi

\bibitem[{{Asensio Ramos} {et~al.}(2007){Asensio Ramos}, {Mart\' inez
  Gonz\'alez}, {L\'opez Ariste}, {Trujillo Bueno}, \& {Collados}}]{andres_07}
{Asensio Ramos}, A., {Mart\' inez Gonz\'alez}, M.~J., {L\'opez Ariste}, A.,
  {Trujillo Bueno}, J., \& {Collados}, M. 2007, ApJ, 659, 829

\bibitem[{{Collados}(1999)}]{manolo99}
{Collados}, M. 1999, in Third Advances in Solar Physics Euroconference, ed.
  B.~Schmieder, A.~Hofmann, \& J.~Staude, 184 (ASP Conference), 3--22

\bibitem[{{Harvey} {et~al.}(2007){Harvey}, {Branston}, \& {Keller}}]{harvey_07}
{Harvey}, J.~W., {Branston}, C.~J., \& {Keller}, C.~U. 2007, ApJ, 177, L180

\bibitem[{{Khomenko} {et~al.}(2003){Khomenko}, {Collados}, {Solanki}, {Lagg},
  \& {Trujillo Bueno}}]{khomenko_03}
{Khomenko}, E.~V., {Collados}, M., {Solanki}, S.~K., {Lagg}, A., \& {Trujillo
  Bueno}, J. 2003, A\&A, 408, 1115

\bibitem[{{Landi degl'Innocenti} \& {Landolfi}(2004)}]{egidio}
{Landi degl'Innocenti}, E. \& {Landolfi}, M. 2004, Polarization in Spectral
  Lines (Kluwer Academic Publishers)

\bibitem[{{Lites}(2002)}]{lites_02}
{Lites}, B.~W. 2002, ApJ, 573, 431

\bibitem[{{Lites} {et~al.}(1999){Lites}, {Rutten}, \& {Berger}}]{lites_99}
{Lites}, B.~W., {Rutten}, R.~J., \& {Berger}, T.~E. 1999, ApJ, 517, 1013

\bibitem[{{Lites} \& {Socas-Navarro}(2004)}]{lites_04}
{Lites}, B.~W. \& {Socas-Navarro}, H. 2004, ApJ, 613, L600

\bibitem[{{Lites} {et~al.}(2007){Lites}, {Socas-Navarro}, {Berger}, {frank},
  {Shine}, {Tarbell}, {Title}, {Ichimoto}, {Katsukawa}, {Tsuneta}, {Suematsu},
  {Shimizu}, \& {Nagata}}]{lites_07}
{Lites}, B.~W., {Socas-Navarro}, H., {Berger}, T., {et~al.} 2007, ApJ, Accepted

\bibitem[{{Livingston} \& {Harvey}(1975)}]{livingston_75}
{Livingston}, W.~C. \& {Harvey}, J.~W. 1975, BAAS, 7, 346

\bibitem[{{Manso Sainz} {et~al.}(2004){Manso Sainz}, {Landi Degl' Innocenti},
  \& {Trujillo Bueno}}]{rafa_04}
{Manso Sainz}, R., {Landi Degl' Innocenti}, E., \& {Trujillo Bueno}, J. 2004,
  ApJ, 614, 89

\bibitem[{{Mart\' inez Gonz\'alez} {et~al.}(2007{\natexlab{a}}){Mart\' inez
  Gonz\'alez}, {Collados}, \& {Ruiz Cobo}}]{marian_spw4}
{Mart\' inez Gonz\'alez}, M.~J., {Collados}, M., \& {Ruiz Cobo}, B.
  2007{\natexlab{a}}, in ASP Conference Series, Vol. 358, 4th Solar
  Polarization Workshop, ed. R.~Casini \& B.~W. Lites, 36

\bibitem[{{Mart\' inez Gonz\'alez} {et~al.}(2007{\natexlab{b}}){Mart\' inez
  Gonz\'alez}, {Collados}, {Ruiz Cobo}, \& {Beck}}]{marian_tesis_07}
{Mart\' inez Gonz\'alez}, M.~J., {Collados}, M., {Ruiz Cobo}, B., \& {Beck}, C.
  2007{\natexlab{b}}, A\&A, submitted

\bibitem[{{Mart\' inez Gonz\'alez} {et~al.}(2007{\natexlab{c}}){Mart\' inez
  Gonz\'alez}, {Collados}, {Ruiz Cobo}, \& {Solanki}}]{marian_07}
{Mart\' inez Gonz\'alez}, M.~J., {Collados}, M., {Ruiz Cobo}, B., \& {Solanki},
  S.~K. 2007{\natexlab{c}}, A\&A, 469, 39

\bibitem[{{Martin}(1987)}]{martin_87}
{Martin}, S.~F. 1987, SoPh, 117, 243

\bibitem[{{Meunier} {et~al.}(1998){Meunier}, {Solanki}, \&
  {Livingston}}]{meunier_98}
{Meunier}, N., {Solanki}, S.~K., \& {Livingston}, W.~C. 1998, A\&A, 331, 771

\bibitem[{{Ram\' irez V\'elez} \& {L\'opez Ariste}(2007)}]{julio07}
{Ram\' irez V\'elez}, J.~C. \& {L\'opez Ariste}, A. 2007, in Memorie della
  Societ\`a Astronomica Italiana, Vol.~78, Solar Magnetism and Dynamics and
  THEMIS Users Meeting, ed. S.~Bommier, V. \& Sahal-Br\'echot, 54

\bibitem[{{S\'anchez Almeida} {et~al.}(2003){S\'anchez Almeida}, {Dom\' inguez
  Cerde\~na}, \& {Kneer}}]{jorge_ita_03}
{S\'anchez Almeida}, J., {Dom\' inguez Cerde\~na}, I., \& {Kneer}, F. 2003,
  ApJ, 597, L177

\bibitem[{{S\'anchez Almeida} \& {Lites}(2000)}]{jorge_00}
{S\'anchez Almeida}, J. \& {Lites}, B.~W. 2000, ApJ, 532, 1215

\bibitem[{{S\'anchez Almeida} \& {Trujillo Bueno}(1999)}]{jorge_99}
{S\'anchez Almeida}, J. \& {Trujillo Bueno}, J. 1999, ApJ, 526, 1013

\bibitem[{{Schlichenmaier} \& {Collados}(2002)}]{schlichenmaier_02}
{Schlichenmaier}, R. \& {Collados}, M. 2002, A\&A, 381, 668

\bibitem[{{Semel}(2003)}]{semel_03}
{Semel}, M. 2003, A\&A, 401, 1

\bibitem[{{Smithson}(1975)}]{smithson_75}
{Smithson}, R.~C. 1975, BAAS, 7, 346

\bibitem[{{Solanki} \& {Stenflo}(1984)}]{sami_84}
{Solanki}, S.~K. \& {Stenflo}, J.~O. 1984, A\&A, 140, 185

\bibitem[{{Stenflo}(1987)}]{stenflo_87}
{Stenflo}, J.~O. 1987, Sol. Phys., 114, 1

\bibitem[{{Trujillo Bueno} {et~al.}(2004){Trujillo Bueno}, {Shchukina}, \&
  {Asensio Ramos}}]{javier_04}
{Trujillo Bueno}, J., {Shchukina}, N., \& {Asensio Ramos}, A. 2004, Nature,
  430, 326

\end{thebibliography}

\end{document}